\documentclass[11pt]{article}

\usepackage{graphics}
\usepackage{multicol}
\usepackage[psamsfonts]{amssymb}

\begin{document}

\thispagestyle{empty}

\small
\begin{center}
\textbf{\large Longitudinal conductivity in Si/SiGe heterostructure \\
at integer filling factors}\\[3mm]
I. Shlimak, V. Ginodman, M. Levin\\
\textit{Minerva Center and Jack and Pearl Resnick Institute of Advanced
Technology,\\
Department of Physics, Bar-Ilan University, Ramat-Gan 52900, Israel}\\[3mm]
M. Potemski, D. K. Maude\\
\textit{High Magnetic Field Laboratory, Max-Planck-Institut f\"{u}r
Festk\"{o}rperforschung/CNRS,\\
F-38042 Grenoble Cedex 9, France}\\[3mm]
K.--J. Friedland\\
\textit{Paul-Drude-Institut f\"{u}r Festk\"{o}rperelektronik,
Hausvogteiplatz 5-7, 10117, Berlin, Germany}\\[3mm]
D. J. Paul\\
\textit{Cavendish Laboratory, University of Cambridge, Madingley Road,
Cambridge CB3 0HE, UK}
\end{center}

\begin{center}
\parbox{132mm}{\footnotesize
We have investigated temperature dependence of the longitudinal conductivity
$\sigma_{xx}$\ at integer filling factors $\nu =i$ for Si/SiGe
heterostructure in the quantum Hall effect regime. It is shown that for odd
$i$, when the Fermi level $E_{F}$ is situated between the valley-split
levels, $\Delta \sigma_{xx}$ is determined by quantum corrections to
conductivity caused by the electron-electron interaction:
$\Delta\sigma_{xx}(T)\sim \ln T$. For even $i$, when $E_{F}$
is located between cyclotron-split levels or spin-split levels,
$\sigma_{xx}\sim \exp[-\Delta_{i}/T]$ for $i=6,10,12$ and
$\sim \exp [-(T_{0i}/T)]^{1/2}$ for $i=4,8$.
For further decrease of $T$, all dependences $\sigma_{xx}(T)$ tend to
almost temperature-independent residual conductivity $\sigma_{i}(0)$. A
possible mechanism for $\sigma_{i}(0)$ is discussed.}
\end{center}

\begin{multicols}{2}

\begin{center}
	\textbf{\footnotesize INTRODUCTION}
\end{center}

The measurement of the temperature dependence of 2D conductivity $\sigma (T)$
in the quantum Hall effect regime is a very useful tool for the analysis of
the density-of-states (DOS) of carriers at different filling factors $\nu $.
At integer filling factors, $\nu =i$ $(i=1,2,3...)$, the Fermi level $E_{F}$
lies in the middle of two Landau levels (LL) where the DOS is minimal and
electron states used to be localized. In this case, the character of
longitudinal conductivity $\sigma _{xx}(T)$ is determined by the ratio
between the energy distance between the two adjacent LLs $E_{i}$ and the
temperature within the measuring interval. If $E_{i}\leqslant T$, one
expects a weak non-exponential dependence for $\sigma _{xx}(T)$, while for $%
E_{i}\gg T$, the conductivity has to be strongly temperature-activated (see,
for example, \cite{Polyakov1,Fogler,Furlan} and references therein),
\begin{equation}
\sigma _{xx}(T)=\sigma _{0}\exp (-\Delta /T).  \label{eq1}
\end{equation}
Here, $\Delta $ is the energy of activation and $2\Delta $ reflects the
mobility gap, which is less than $E_{i}$ because of the non-zero width of
the band of delocalized states in the center of each LL, the prefactor $%
\sigma _{0}$ is equal to $2e^{2}/h$~\cite{Polyakov1}. (The coefficient 2
appears because the conductivity is due to electrons excited into the upper
LL and holes in the lower LL.) For large $\Delta \gg T$, direct excitations
of electrons to the mobility edge is unlikely and the conductivity is due to
the variable-range-hopping (VRH) mechanism via localized states in the
vicinity of $E_{F}$:~\cite{Polyakov2,Haug,Shin}
\begin{equation}
\sigma _{xx}(T)\propto \exp (-T_{0}/T)^{m},  \label{eq2}
\end{equation}
where $m=1/2$ because of the existence of a Coulomb gap in the DOS at
$E_{F}$ \cite{Efros,the Book}. The parameter $T_{0}$ is connected with the
localization radius $\xi \left( \nu \right) $ of the states for given $\nu $:
$T_{0}=C_{1}e^{2}/\kappa \xi (\nu) $. Here $C_{1}\approx 6$
for two dimensions and $\kappa $ is the dielectric constant of the host
semiconductor.

Most previous measurements of $\sigma _{xx}(T)$ were performed on
GaAs/AlGaAs heterojunctions. Increased interest in the study of the Si/SiGe
heterostructure is motivated by the application of thin Si$_{1-x}$Ge$_{x}$
layers as the base of a heterojunction bipolar transistor with increased
mobility \cite{Cressler}, resonant interband tunneling diodes \cite{Rommel},
as well as by possible future application of this heterostructure for
quantum computing \cite{Vrijen,Shlimak}. The special feature of $n$-type
Si/SiGe in comparison with GaAs/AlGaAs heterostructures lies in the
appearance of an additional splitting of energy levels due to lifting of
two-fold valley degeneracy in a strong perpendicular magnetic field. As a
result, in $n$-type Si/SiGe heterostructures, odd filling factors correspond
to the location of $E_{F}$ between valley-splitting LLs.

In Ref.~\cite{Weitz} measurements of $\sigma _{xx}(T)$ in tilted magnetic
fields were used for determining the valley-splittings in
Si/Si$_{1-x}$Ge$_{x}$ heterostrucuture. It was found that the values of
$\Delta $ for odd $i=3,5,7,9$, as determined from the Arrhenius plot, Eq.~(1),
do not agree with the
\linebreak

\vspace{-5mm}
\begin{center}
\setlength{\unitlength}{1mm}
\begin{picture}(85,53)
	\includegraphics[85mm,58mm]{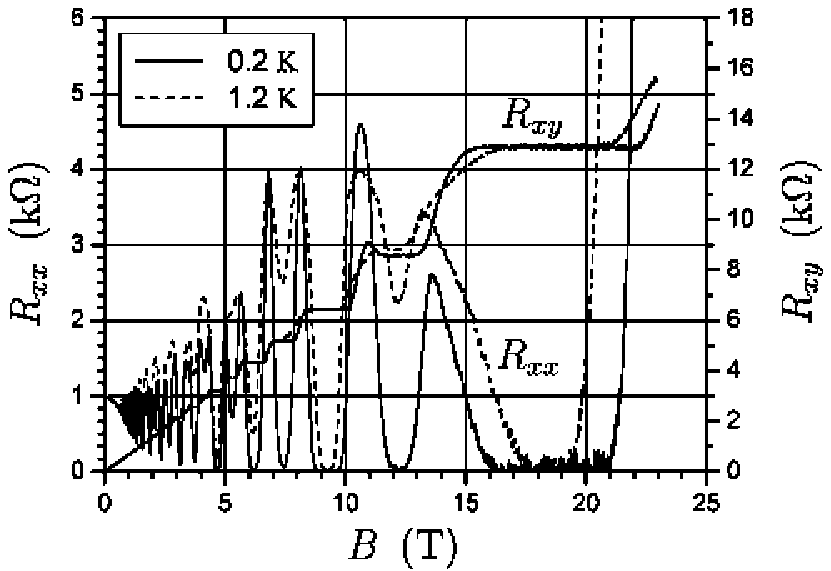}
\end{picture}
\end{center}
{\footnotesize FIG. 1:
Transverse resistance $R_{xy}$ and longitudinal resistance $R_{xx}$
of Si/Si$_{0.7}$Ge$_{0.3}$ heterostructure at $T$ = 0.2~K and 1.2~K.}
\vspace{5mm}

\noindent
values of $E_{i}$ estimated theoretically in
Ref.~\cite{Ando,Ohkawa}.
However, the values obtained for $\Delta $ (0.2--0.4~K) were of order $T$,
which makes doubtful the use of the Arrhenius law for data processing. In
the same work \cite{Weitz}, the coincidence method in tilted magnetic fields
was used to determine spin-splitting and the effective $g$-factor $g^{\ast }$.
It was found that $g^{\ast }=3.4$ for filling factors $16\leq \nu \leq 28$
and increases for lower $\nu $. The spin- and valley-split energy levels
were also determined in strained Si quantum wells using Shubnikov-de Haas
oscillations measurements \cite{Koester}. It was found that for a
perpendicular magnetic field of $\sim 2.8$ T wich corresponds to $\nu =7,$ a
valley splitting is of order of 52 $\mu $eV $\approx 0.6$ K. This value is
in agreement with the data obtained in \cite{Weitz} for
Si/Si$_{1-x}$Ge$_{x}$ heterostrucuture, but is much less than
those determined for strained inverse layer in Si-MOS structures in strong
magnetic fields: $\Delta \approx 12$~K for $B=14.6$~T, Ref.~\cite{Nicholas} or
\linebreak
$\Delta $ [K]$=2.4+0.6\cdot B$ [T] at 2T
\mbox{$<$}$B$\mbox{$<$}
8T, Ref.~\cite{Pudalov-1}. It was also shown in Ref.~\cite{Koester} that
$g^{\ast }\approx 3.5$ at $\nu \geq 10$ and $g^{\ast }$ oscillates between
2.6 and 4.2 with decreasing $\nu $. To summarize, the character of
$\sigma_{xx}(T)$ for different $i$ in Si/SiGe heterostructure remains vague,
which motivated this work.
\vspace{2mm}

\begin{center}
	\textbf{\footnotesize EXPERIMENT}
\end{center}
\vspace{1mm}

The sample investigated was Hall-bar patterned n-type Si/Si$_{0.7}$Ge$_{0.3}$
double heterostructure, 7 nm i-Si quantum well was situated between 1~$\mu$m
i-Si$_{0.7}$Ge$_{0.3}$ layer and 67~nm Si$_{0.7}$Ge$_{0.3}$ layer with 17~nm
spacer followed by 50~nm Si$_{0.7}$Ge$_{0.3}$ heavily doped with As. A 4~nm
silicon cap layer protects the surface. The
\linebreak

\vspace{-5mm}
\begin{center}
\setlength{\unitlength}{1mm}
\begin{picture}(85,67)
	\includegraphics[85mm,65mm]{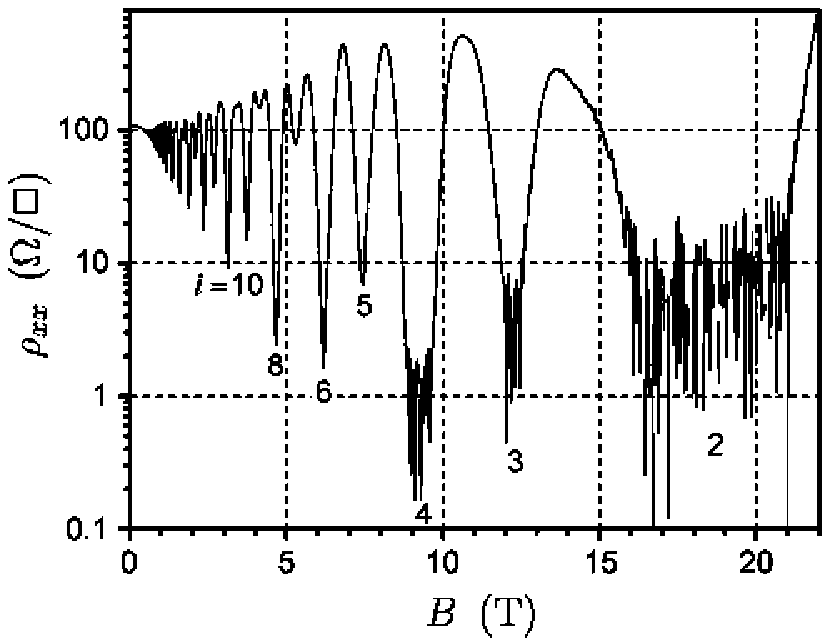}
\end{picture}
\end{center}
\vspace{-3mm}
{\footnotesize FIG. 2:
Longitudinal resistivity $\rho _{xx}=R_{xx}/\square $ on
logarithmic scale at $T=0.2$ K. The values of $i$ are shown near the
minima.}
\vspace{5mm}

\noindent
electron concentration $n$ and mobility $\mu $ at 1.5~K were
$n=9\cdot 10^{11}$~cm$^{-2}$, $\mu =$ 80,000~cm$^{2}$/V$\cdot $s.
The sample  resistance was measured using a standard lock-in-technique,
with the measuring current being 20~nA at a frequency of 10.6~Hz.

Figure 1 shows the longitudinal $R_{xx}$ and transverse $R_{xy}$ resistances
of the sample investigated when measured at $T=1.2$ and 0.2~K in magnetic
fields up to $B=$ 23~T. The plateau in $R_{xy}$ are clearly seen at values
which are a portion of a quantized resistance $h/e^{2}=25.8$~k$\Omega $. At
some magnetic fields $B_{i}$ when the filling factor $\nu $ achieves an
integer value $\nu =i$, longitudinal resistance $R_{xx}$ exhibits a deep
minimum, these fields correspond approximately to the midpoint of each
$R_{xy}$ plateau.

Figure 2 shows the two-dimensional resistivity $\rho _{xx}=R_{xx}/\square $
on a logarithmic scale for $T=0.2$~K. At $\nu $ around $i=2,3,4$, huge
fluctuations of $\rho _{xx}$ are seen. These fluctuations of longitudinal
voltage $\Delta V_{xx}$ do not reflect fluctuations of the sample
resistivity or sample inhomogeneity, but can be explained by the fact that
in strong magnetic fields and small integers $i$, both 2D resistivity
$\rho_{xx}$ and conductivity
$\sigma _{xx}=\rho _{xx}/\left( \rho _{xx}^{2}+\rho_{xy}^{2}\right) $
are close to zero, which leads to decoupling of the bulk
of 2D electron system from the contacts at the edges \cite{Weis}. These
fluctuations are also not
connected with the scan rate of the magnetic
field, because they were observed in experiments when the magnetic field is
fixed and only temperature is variable (Fig.~3). The magnitude of
$\Delta V_{xx}$ increases with decreasing $\nu $: for $\nu $ around $i=4,3,2$,
the maximal values of $\Delta V_{xx}$ achieved to 0.5~$\mu $V, 1~$\mu $V and
4~$\mu $V correspondingly. These fluctuations of the voltage
signal prevent
from determination of $\sigma _{xx}$\ at $i=2$. For
\linebreak

\vspace*{-5mm}
\begin{center}
\setlength{\unitlength}{1mm}
\begin{picture}(85,65)
	\includegraphics[85mm,65mm]{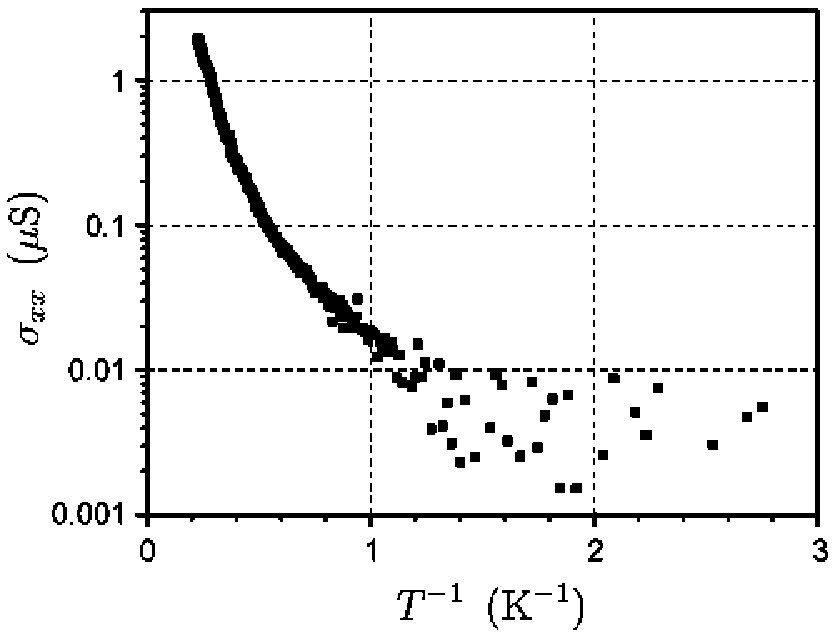}
\end{picture}	
\end{center}
{\footnotesize FIG. 3:
$\sigma _{xx}(T)$ for $\nu =4$ for magnetic field fixed at
$B_{4}=9.1$~T.}
\vspace{5mm}

\noindent
the same reason, we will
not discuss $\sigma _{xx}(T)$ for $i=3,4$ below $T=0.2$~K.\medskip

\begin{center}
	\textbf{\footnotesize RESULTS AND DISCUSSION}
\end{center}

\textit{Odd integers} ($i=3,5,7,9$).

In the case of $n$-Si/SiGe heterostructure, odd filling factors correspond
to the location of $E_{F}$ between the valley-split
LLs. The valley
splitting of strained Si layers has been theoretically investigated in
Ref.~\cite{Ando,Ohkawa,Koeler}. It was shown in Ref.~\cite{Ohkawa} that valley
splitting could be observed only in the presence of a high magnetic field
normal to the interface and is given approximately by
\begin{equation}
\varepsilon _{v} [\mbox{K}]\approx 0.174\cdot (N+1/2)\cdot B[\mbox{T}].
\label{eq3}
\end{equation}

Here the valley splitting energy $\varepsilon _{v}$ is measured in Kelvin,
magnetic field $B$ in Tesla, $N=0,1,2...$ is the Landau index. Because
increase of $B$ is accompanied by decrease of $N,$ the values of the
valley-splitting weakly depend on $B$. Numerical estimation based on Eq.~(3)
showed that the values of $\varepsilon _{v}$ for magnetic fields 4T
\mbox{$<$}$B$\mbox{$<$}12T are about 1~K.
Therefore, one cannot expect an activated character of
$\sigma _{xx}(T)$ within the experimental temperature interval
($T=$4.2--0.2~K).

In contrast, in \cite{Pudalov-1} much larger values of $\varepsilon _{v}$
have been reported with a significant energy of the valley-splitting even
without magnetic field, it was emphasized that these data agree well with
theoretical calculations \cite{Koeler}:
\begin{equation}
\varepsilon _{v}[\mbox{K}]\approx 2.4+0.6\cdot B[\mbox{T}].  \label{eq4}
\end{equation}

In accordance with Eq. (4), the value of $\varepsilon _{v}$ for $i=3$ in the
case of our sample ($B\approx 12$T) should be about
\linebreak

\vspace*{-5mm}
\begin{center}
\setlength{\unitlength}{1mm}
\begin{picture}(85,65)
	\includegraphics[85mm,65.3mm]{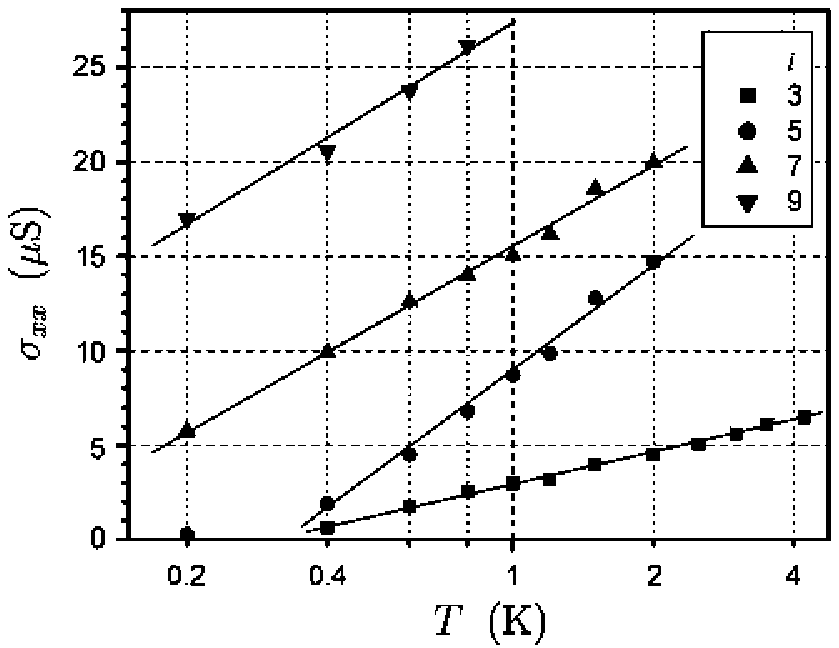}
\end{picture}
\end{center}
\vspace{-3mm}
{\footnotesize FIG. 4:
$\sigma _{xx}$ for odd integers $i=3,5,7,9$ as a function of
$\ln T$.}
\vspace{5mm}

\noindent
10 K providing strong
activated character of conductivity. However, the analysis shows that
$\Delta \sigma _{xx}$ weakly depends on $T$, there is no activation process,
the best fit of experimental data is achieved by logaritmic law:
$\Delta \sigma _{xx}(T)\sim \ln T$ (Fig.~4). This result agrees
with the model of Ohkawa and Uemura \cite{Ohkawa}.

Logarithmic temperature dependence of conductivity at low temperatures is
usually interpreted as a manifestation of corrections to the conductance due
to quantum interference effects \cite{Altshuler,Lee}. In strong
perpendicular magnetic fields, weak localization corrections to the
conductivity are suppressed and $\Delta \sigma _{xx}$ is determined by
quantum corrections due to the electron-electron interaction, which occurs
both in weak and in strong magnetic fields (see, for example,
Ref.~\cite{Murzin} and references therein).
This leads to the following expression for
the temperature correction to the conductivity \cite{Hikami}:
\begin{equation}
\Delta \sigma _{ee}(T)=\left( \frac{\alpha pe^{2}}{2\pi h}\right) \ln \left(
\frac{T}{T_{ee}}\right),  \label{eq5}
\end{equation}%
where $\alpha $ is a constant of order unity and $p$ is the exponent in the
temperature dependence of the phase-breaking time
$\tau _{\varphi }\sim T^{-p}$. At low $T$, the phase used to be
broken by the electron-electron
interaction, leading to $p\approx 1$ \cite{Altshuler}. This gives
\begin{equation}
(1/\alpha )\frac{\Delta \sigma _{xx}}{\left( e^{2}/h\right) }=\frac{1}{2\pi }%
\ln \left( \frac{T}{T_{ee}}\right).  \label{eq6}
\end{equation}

In Fig. 5, the dimensionless conductivity is plotted as a function of
dimensionless temperature $T/T_{ee}$, the values of $T_{ee}$ being
determined from the intersection with the $x$-axis for each curve in Fig.~4.
The solid
\linebreak

\vspace*{-5mm}
\begin{center}
\setlength{\unitlength}{1mm}
\begin{picture}(85,64)
	\includegraphics[85mmm,64.3mm]{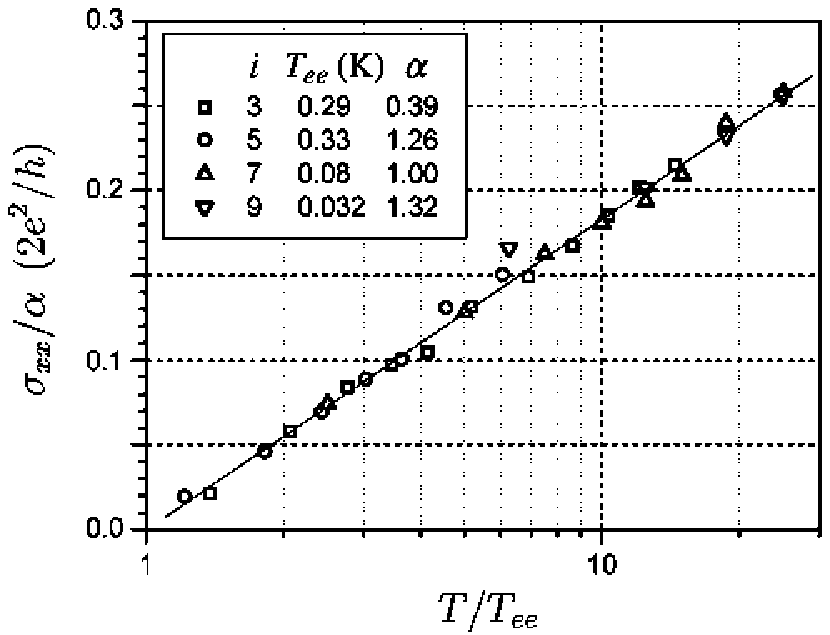}
\end{picture}
\end{center}
{\footnotesize FIG. 5:
$\sigma _{xx}(T)/\alpha $ for odd integers plotted in
dimensionless units $\sigma _{xx}/\alpha (2e^{2}/h)$ vs. $\ln (T/T_{ee})$.
The insert shows the values of the adjustable parameter $\alpha $ and
$T_{ee}$.}
\vspace{5mm minus 2mm}

\noindent
line in Fig.~5 corresponds to the slope $(1/2\pi )$. Having
$\alpha $ as the only adjustable parameter, one can merge all curves. The
insert shows the values obtained for $\alpha $ which are indeed of order
unity. Thus, $\sigma _{xx}(T)$ for odd integers can be successfully
described in terms of quantum corrections to the conductivity in strong
magnetic fields caused by the electron-electron interaction.\medskip

\textit{Even integers} ($i=4,6,8,10,12$).

For even integers, there are two possibilities for the location of $E_{F}$:
between cyclotron LLs (four-multiple integers $i=4,8,12$) and between
spin-split levels $(i=6,10)$. Taking into account that for the strained Si
well, $m^{\ast }=0.195m_{0}$ \cite{Murphy}, the cyclotron energy is given by:
\begin{equation}
\hbar \omega _{C}[\mbox{K}]=6.86B[\mbox{T}].  \label{eq7}
\end{equation}

The energy of spin splitting $g^{\ast }\mu _{B}B$ depends on the effective $%
g $-factor $g^{\ast }$. As mentioned earlier, the value of $g^{\ast }$
increases for lower $\nu $ oscillating between 2.6 and 4.2~\cite{Koester}.
For numerical estimates, we assume $g^{\ast }\approx 3.8$, giving
\begin{equation}
g^{\ast }\mu _{B}B[\mbox{K}]\approx 2.55B[\mbox{T}].  \label{eq8}
\end{equation}

In Ref.~\cite{Pudalov}, a similar value ($g^{\ast }\mu _{B}B\approx $
2.6~K/T) was used for estimating spin-splitting in Si-inversion layers in
high-mobility Si-MOSFETs. In the calculation of $E_{i}$, all relevant
splitting energies are taken into account. For example,
$E_{4}=\hbar \omega_{C}-g^{\ast }\mu _{B}B-\frac{1}{2}[\varepsilon
_{V}^{N=0}+\varepsilon _{V}^{N=1}]$.
Substituting $B_{4}=9.1$~T in Eqs.~(3),
(7), (8), we obtain $E_{4}\approx 40$~K. Similarly,
$E_{6}=g^{\ast }\mu_{B}B-\varepsilon _{V}^{N=1}$.
Substituting $B_{6}=6.07$~T in Eqs.~(3) and
(8), one get $E_{6}\approx 12$~K and so on. These energies are shown in the
insert in Fig. 6. Because all $E_{i}$ are larger than $T$ within the
experimental interval of temperatures, it is expected that
\linebreak

\vspace{-5mm}
\begin{center}
\setlength{\unitlength}{1mm}
\begin{picture}(85,69)
	\includegraphics[85mm,68.7mm]{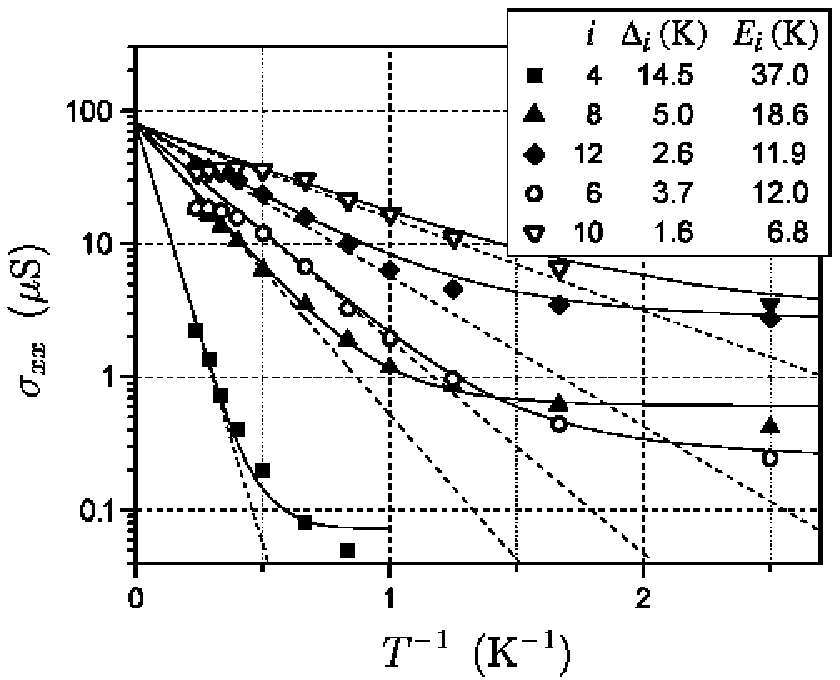}
\end{picture}
\end{center}
{\footnotesize FIG. 6:
$\sigma _{xx}$ for even integers $i=4,8,12$ and $6,10$,
plotted on Arrenius scale of $\ln \sigma _{xx}$ vs. $1/T$. The insert shows
the values of the experimentally determined $\Delta _{i}$ and the calculated
values of $E_{i}$. Solid curves represent the calculated dependences
$\sigma_{xx}(T)=(2e^{2}/h)\exp \left( -\Delta _{i}/T\right) +\sigma _{i}(0),$ with
$\sigma _{i}(0)$ as the only adjustable parameter. The values of
$\sigma_{i}(0)$ are shown in insert in Fig.~7.}
\vspace{5mm}

\noindent
$\sigma _{xx}(T)$
will be determined by the temperature-activated excitation of electrons to
the mobility edge and characterized, therefore, by the constant energy of
activation $\Delta \lesssim 1/2E_{i}$, Eq.~(1). In Fig.~6, the dependences
$\sigma _{xx}(T)$ for even integers are plotted on this scale,
$\ln \sigma_{xx}$ vs. $1/T$. One sees, that at high temperatures,
the experimental
points are in agreement with Eq.~(1), which allows to determine the values
of $\Delta _{i}.$ The prefactor $\sigma _{0}$ for all curves is close to
$78\mu $S $\approx 2e^{2}/h$, in accordance with the theoretical
prediction \cite{Polyakov1}. It is seen, however, that with decrease of
temperature, all dependences tend towards the residual, almost
temperature-independent conductivity $\sigma _{i}(0)$. Having
$\sigma_{i}(0) $ as the only adjustable parameter in expression
$\sigma_{xx}(T)=(2e^{2}/h)\exp \left( -\Delta _{i}/T\right) +\sigma _{i}(0)$,
the values of $\sigma _{i}(0)$ were determined from fitting the calculated
$\sigma _{xx}$ (solid lines in Fig.~6) to the experimental points.
Subtraction of $\sigma _{i}(0)$ allows us to merge all curves into one
straight line on the dimensionless scale,
$\ln \{[\sigma _{xx}(T)-\sigma_{i}(0)]/(2e^{2}/h)\}$ vs.
$\Delta _{i}/T$ (Fig.~7).

It follows from Figs. 6 and 7 that the low-temperature experimental points
for $i=4$ and 8 do not fit well to the
calculated curves. This can be
explained by the fact that the values of $\Delta _{i}$ for $i=$ 4 and 8 are
substantially larger than for $i=$ 6, 10, 12. As a result, direct thermal
excitation of electrons to the mobility edge is unlikely and it is more
probably that electron transport is due to variable-range-hopping (VRH)
conductivity via localized states in the vicinity of $E_{F}$,
\linebreak

\vspace*{-5mm}
\begin{center}
\setlength{\unitlength}{1mm}
\begin{picture}(85,68)
	\includegraphics[85mm,66mm]{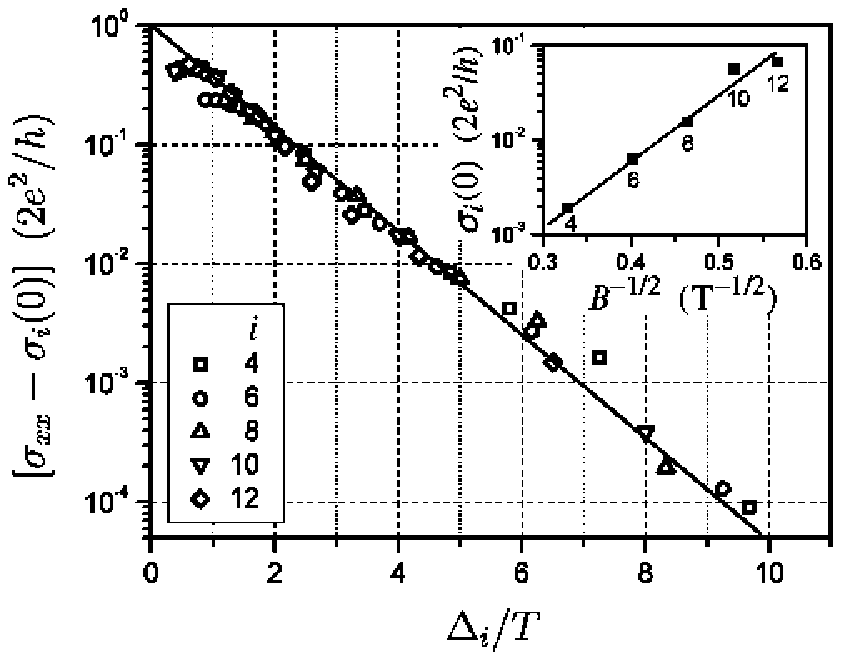}
\end{picture}
\end{center}
{\footnotesize FIG. 7:
$\sigma _{xx}(T)-\sigma _{i}(0)$ for even integers in
dimensionless units as a function of a dimensionless reciprocal temperature
$\Delta _{i}/T$. The insert shows the residual conductivity $\sigma _{i}(0)$
in units of $(2e^{2}/h)$ as a function of magnetic fields $B$ for different
even $i$ shown near the points, with the straight line corresponding to
$\sigma _{i}(0)\propto \exp (B^{-1/2})$.}
\vspace{5mm}

\noindent
Eq.~(2). To
summarize, one can write the general expression for the longitudinal
conductivity at even filling factors:
\begin{equation}
\sigma _{xx}(T)=\frac{2e^{2}}{h}\exp (-\frac{\Delta _{i}}{T})+\sigma _{0}\exp
(-\frac{T_{0i}}{T})^{1/2},  \label{eq9}
\end{equation}
where the first term corresponds to activation of localized electrons from
the Fermi level $E_{F}$ to the mobility edge, while the second term
corresponds to VRH in the vicinity of $E_{F}$. If $\Delta _{i}\gg T$, the
first term is very small and the second term dominates in $\sigma _{xx}$.

To check this assumption, we plot $\sigma _{xx}(T)$ for $i=$ 4 and 8 in the
VRH scale of Eq.~(2): $\ln \sigma _{xx}$ vs. $T^{-1/2}$ (Fig.~8). On this
scale, all experimental points for $i=4,8$ coincide with the calculated
curves $\sigma _{xx}(T)=\sigma _{0}\exp (-T_{i0}/T)^{1/2}+\sigma _{i}(0)$,
where $\sigma _{0}$ and $T_{i0}$ are determined from experiment and
$\sigma_{i}(0)$ is the only adjustable parameter. (As expected, experimental
data $\sigma _{xx}(T)$ for $i=$ 6, 10 and 12 do not fit well to the VRH scale
and therefore are not shown in Fig.~8). The insert in Fig.~7 shows
$\sigma_{i}(0)$ obtained for different $i$ as a function of magnetic field
$B_{i}$.
It is found that the best fit corresponds to the exponentially strong
dependence $\sigma _{i}(0)\propto \exp (B_{i}^{-1/2})$.

The question arises about the origin of residual conductivity
$\sigma_{i}\left( 0\right) $. It worth to emphasize that the
low-temperature
saturation of longitudinal conductivity (or resistivity) in the quantum Hall
effect regime is not a new phenomena, it had been observed earlier
\linebreak

\vspace*{-5mm}
\begin{center}
\setlength{\unitlength}{1mm}
\begin{picture}(85,70)
	\includegraphics[85mm,67.5mm]{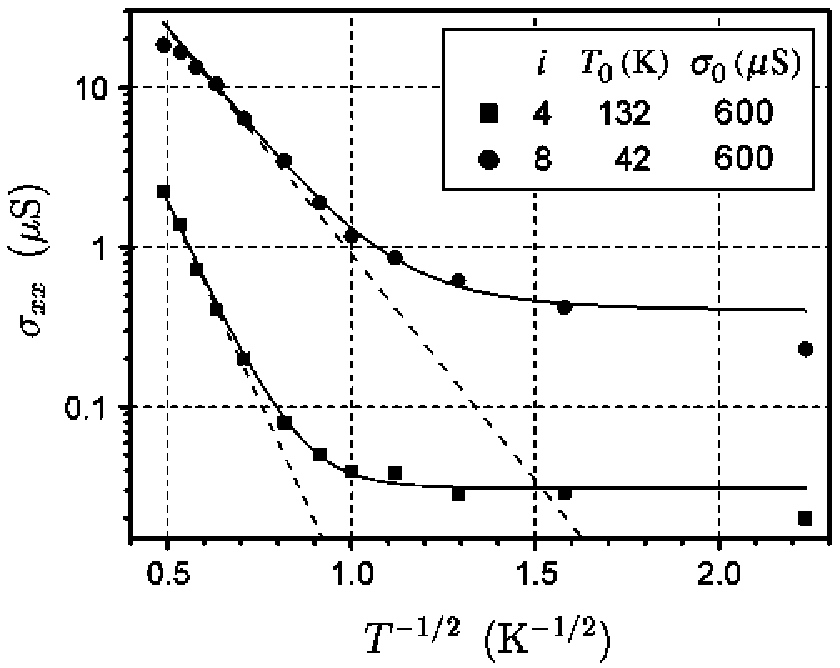}
\end{picture}
\end{center}
{\footnotesize FIG. 8:
$\sigma _{xx}(T)$ for $i=4,8$ plotted in the VRH scale $\ln\sigma _{xx}$
vs. $T^{-1/2}$. Solid curves correspond to the calculated dependences
$\sigma _{xx}(T)=\sigma _{0}\exp (-T_{i0}/T)^{1/2}+\sigma_{i}(0)$,
with the values of $T_{i0}$ and $\sigma _{0}$ shown in the
insert. The values of $\sigma _{i}(0)$ are shown in the insert in Fig.~7.}
\vspace{5mm}

\noindent
in modulated doped GaAs/AlGaAs~\cite{Furlan} and Si/SiGe~\cite{Weitz}
heterostructures. However, we are not aware of any discussion of the origin
of this effect. Let us enumerate the experimental features of the residual
conductivity: (i) The values of $\sigma _{i}(0) $ are much
smaller than the minimal quantum for 2D conductivity
$e^{2}/h\approx 39$~$\mu$S;
\linebreak
(ii) $\sigma _{i}(0) $
decreases strongly with increasing magnetic field
$B: \sigma _{i}(0)\propto \exp (B_{i}^{-1/2})$;
(iii) $\sigma _{i}(0) $ exists in all investigated
temperatures, which means that this mechanism of conductivity occurs in a
parallel conductive channel.

Both (i) and (ii) suggest that $\sigma _{i}(0)$ is a sort of hopping
conductivity. Indeed, $e^{2}/h$ is the minimal value of metallic
conductivity in 2D, while $\sigma _{i}(0)\ll e^{2}/h$. Moreover, there is no
mechanism of exponentially strong magnetoresistance for metallic
conductivity. By contrast, in strong magnetic fields, hopping resistivity
$\rho _{3}$ increases exponentially:
$\rho _{3}\propto \exp (\mbox{const}\cdot B_{i}^{1/2})$~\cite{the Book}.
However, weak temperature dependence of
$\sigma _{i}(0)$\ contradicts to the hopping model and needs additional
assumptions. We believe that this can be explained by the non-equilibrium
character of $\sigma _{i}(0)$, which means the absence of thermal
equilibrium in the distribution of electrons across the localized states, as
was observed earlier in electron glasses \cite{Vaknin}. A very slow rate of
relaxation can be caused, for example, by an exponential decay of the DOS in
the vicinity of $E_{F}$. In this case, relaxation to the lower states with
decreasing temperature requires hopping over long distances and therefore is
very
\linebreak
unlikely. If the regions with such modified DOS form a continuous path
along the voltage probes, a parallel conductive channel will appear, which
explains (iii). At low temperatures, a weakly temperature-dependent residual
conductivity will override the activated conductivity of the bulk 2D
plane.
\vspace{10mm}

\begin{center}
	\textbf{\footnotesize ACKNOWLEDGMENTS}
\end{center}

We are thankful to B.~I.~Shklovskii for discussion, A.~Belostotsky for the
help with data analysis, and to the Eric and Sheila Samson Chair of
Semiconductor Technology for financial support. V.~G. and M.~L. thank the
``KAMEA'' Programme for support.

\end{multicols}
\vspace{3mm}

\begin{center}
	\rule[2.5mm]{120mm}{0.2mm}\\
	\vspace{-\baselineskip}
	\rule[2.4mm]{80mm}{0.4mm}\\
	\vspace{-\baselineskip}
	\rule[2.3mm]{40mm}{0.6mm}
\end{center}

\begin{multicols}{2}

\end{multicols}

\end{document}